Matias Slavov, Postdoc at LUT-University, Lappeenranta

# Timekeeping Beyond Human Whim: A Critical Analysis of Strong Conventionalism about Clocks

***Abstract:*** Conventional elements play an integral role in measuring time. Clocks unavoidably require a chosen standard of synchrony. The periodic process selected for the standard is not imposed upon us in any way. Conventionalism, in this regard, offers an insightful perspective on temporal measurement. The strong formulation of this doctrine is, however, questionable. After elaborating on strong conventionalism, it will be argued that it has three downsides: (i) it fails to properly distinguish between highly stable natural processes and complete randomness, as periodicities differ in terms of regularity; (ii) it does not explain that a clock actually measures something, to wit, parts of its own path along its worldline in spacetime; (iii) and relatedly, it omits the distinction between a temporal interval as a physical quantity and the arbitrarily chosen unit of time, leaving the distinction between temporal quantities and units a mystery. This article does not claim that strong conventionalism is incorrect; rather, considering the critical arguments against it, we may conclude that it remains an incomplete theory of what clocks measure.

## 1. Introduction

Conventionalism has multiple fields of applicability. One might be a conventionalist about property, geometry, and necessity, among others (Rescorla 2019). To provide a cursory definition, something is a convention if it is not forced upon us, but it is in some sense based on our choice. This much is not controversial in the characterization of what it is to be a convention. Typically, conventionalists aim to disclose that what we take to be natural is due to a social agreement. *The Oxford Dictionary of Philosophy* specifies that conventionalism "magnifies the role of decisions, or free selection from amongst equally possible alternatives, in order to show that what appears to be objective or fixed by nature is in fact an artefact of human convention" (Blackburn 1996, 81–2). The central difficulty is in assessing whether all choices made by humans are as good. Are there better or worse conventions, or are all of them merely arbitrary? Could one convention somehow track a real natural property, an aspect of the world that sustains itself independently of any human involvement?

Regarding time, the conventional status of simultaneity and direction have been discussed extensively.[1] No consensus – at least on all specific, intricate details – has been reached on these matters, and this article will not cover these concepts. The focus shall be on conventionalism about clocks. Although somewhat conventional remarks can be found already in figures like Hume and Mach, Carnap is the most famous proponent of conventionalism about clock measurement.

[1] Conventionalism about spacelike separation maintains there is no fact of the matter about the temporal order of such events, not even in a specific inertial frame of reference, as the order depends on the choice of the synchronization parameter. Conventionalism about direction indicates that forward and backward directions about time describe the world equally well. It is only the increase of entropy that we use to distinguish between the two directions. Yet the second law of thermodynamics concerns the macroscopic realm. The direction of time might be due to our blurred vision: if we could see fundamental microscopic processes, we could not tell time's direction. An overview of conventionalism about simultaneity is provided by Janis (2018). For a recent defense of conventionality of direction, see Farr (2022). Rovelli (2023) has recently explored the foundation of causal asymmetry in thermodynamics.

Carnap's position has been criticized by Tal (2016). The account of Tal remains "metaphysically neutral" as it neither presupposes nor denies "any sort of realism about measureable quantities" (Tal 2016, 310). This article, for its part, takes a different route as it tackles conventionalism from a metaphysically realist viewpoint. Clocks are instruments that measure a real mind-independent quantity, parts of the interval of their own timelike worldlines. In this sense, clocks are similar kinds of devices to scales or thermometers, as they track definite quantities of mass (by measuring weight) and the average kinetic energy of the system. Strong conventionalism omits the fact that measuring something successfully is about measuring something about objective reality.

Before getting into the issue of temporal conventionalism, we should be clear about what the relevant notion of a clock is in this context, and what the core philosophical issue about clock accuracy is (Section 2). Analyzing the definition of a clock and the process of how clocks get synchronized leads us directly to the issue of conventionalism (Section 3). Although conventionalism has significant virtues in assessing temporal measurement, the strong version of this doctrine shall be analyzed critically (Section 4). Although this article will not prove or try to prove strong conventionalism wrong, it will be argued that there are three defects to strong conventionalism: it does not adequately consider different degrees of regularity among periodic processes (Section 4.1); it omits the theoretical-empirical notion of accuracy as the proper time measured by a clock along its worldline (Section 4.2); it leaves the distinction between temporal units and temporal quantities a mystery (Section 4.3). The conclusion is that, despite there being unavoidably conventional elements about clocks, strong conventionalism is not a credible view about temporal measurement (Section 5).

## 2. The definition of a clock

A clock is involved in a repetitive process. A good clock is thought to instantiate regularity. One circle of time is like another in duration. These measured durations are ideally congruent. A clock has a mechanism in counting ticks, which also should ideally be congruent. It must have a way for converting the counted ticks to display of time. Many naturally occurring cycles, such as spinning and orbits of planets/moons, as well as animal circadian rhythms, are often linked to the functioning of clocks, but as they include neither counting nor display, they are not clocks. If clocks, by definition, relate to cyclical processes, then stopwatches and timers are not properly clocks either. They measure durations and count time down, but there is no regular period that repeats over and over (Dowden 2025).

A clock needs to be synchronized with the aid of some repetitive physical process. There are great many standards of synchronization humans have used throughout history. Since the 1960s, the official standard of synchrony has been based on the periodic vibration rates of cesium atoms, on the behavior of its quantum mechanical spins. Second has been officially defined "to be the duration of 9,192,631,770 periods of radiation corresponding to the transition between the two hyperfine levels of the ground state of the cesium 133 atom" (Gibbs 2002, 87). The practical advantages of such atomic clocks as a standard for global timekeeping are the i) similar behavior of all cesium-133 atoms, ii) the regular ticking of the clock based on these atoms, iii) high frequency as a basis of the ticks, and iv) the ability to copy and construct a similar clock elsewhere. The current cesium-based clocks are arguably not the best available. They operate with microwave frequencies, whereas there are optical clocks that tick at higher frequencies and reportedly are about 100 times more accurate (Johnston 2021).

Reports on the science and engineering of clocks typically assume that one clock is more accurate than another. The underlying question – What is the basis of clock accuracy? – is left unanswered. Pre-relativistically, the Newtonian argument gives a straightforward answer. If there is an independent dynamic structure of time, that is, a time itself that flows equably, then there is a completely objective standard for the subsequent ticks of a clock or its periods to be isochronous. The standard for accuracy is provided by the equal intervals of the uniformly passing time. When clock smiths and engineers adjust clocks, they do not adjust them only in reference to other clocks, or some measurable and observable cyclical processes. The Newtonian argument suggests that the more accurate a clock is, the better it imitates true temporal intervals of time itself. For the ticks to occur at equal intervals and for the pointer to circle around at a constant rate, the ticks must be equal and the sweepings constant with respect to the passage of a unique, space-independent self-existing time (Maudlin 2012, 15–6).

Under the orthodox relativistic account of the world, there is no unique passage of time. Due to the conventionality and relativity of simultaneity, there is no universal present moment stretched throughout the entire universe. Gödel (1949) already remarked that there is no unique universal hyperplane of simultaneity and a constant new creation of the present moment. According to common wisdom, relativity is at odds with presentism — roughly, the view that only the present time and presently existing entities exist — and in favor of eternalism — roughly, the view that all times, past, present and future, and all entities exist unconditionally.[2] If eternalism is true, there cannot be any objective change along past, present and future. All times simply exist, and it is a perspectival matter as to what counts as past, present or future. My main aim here is not to discuss the dialectic between presentism and eternalism. For this article the essential point is that there are good reasons to question the existence of a unique robust passage of time in modern relativistic physics.[3]

---

[2] The literature on the implications of relativity to the metaphysics of time is voluminous. To just list a few 21st century contributions, Saunders (2002), Balashov and Janssen (2003), and Peterson and Silberstein (2010) think that relativity is clearly against presentism. Craig (2001), Swinburne (2008), and Amiriara (2021) disagree.

[3] Some have argued for deflationary theories in which passage is defined as a succession of events along observers' worldlines. This relativistic account allows there to be multiple

From a relativistic perspective, it is questionable if there is a completely unobservable, non-testable, and non-measurable substantial time. Such a Newtonian time includes absolute simultaneity, a privileged foliation of spacetime. All events across the entire universe and everything that physically exists should be connected with a hyperplane of simultaneity, and their temporal difference should be uniquely zero. All successive events, for their part, should have a unique positive temporal difference. This is evidently in contradiction with both the relativity and the conventionality of simultaneity. Moreover, Einsteinian relativity makes time a physical and empirical phenomenon: simultaneity and succession have to do with clocks and electromagnetic signals traversing between them.[4]

In lack of absolute standard for synchrony, we must deal with conventionally chosen cyclical processes for the construction and adjustment of clocks. It is obvious there is unavoidably human choice involved. This motivates conventionalism about temporal measurement.

## 3. Strong conventionalism about clocks

Conventionalism, at least in its most rigid form, can be found from Carnap's measurement theory. According to his strong conventionalism concerning temporal measurement, any choice of standard is equally valid. There is no sense to ask whether some periodic processes themselves exhibit congruity because those processes themselves define congruity. Denying that there is an absolute standard for the synchronization of clocks is not a new idea. Hume, for one, already argued that the standards of synchrony are relative (Slavov 2021, Section 3.2). In

---

passages of time, but no one unique passage. For example, Fazekas (2016), Deng (2019), Leininger (2021) and Slavov (2022) consider something like a deflationary account. Moreover, the moving spotlight theory is eternalist in the sense it predicates the existence of all times but maintains the present tense to be absolute. It therefore includes robust passage.

[4] This is apparent already in Einstein's (1905) original contribution on the electromagnetism of moving bodies.

his *Treatise of Human Nature*, he articulates his stance on time-measurement briefly. The relevant parts (T 1.2.4.24; SBN 47–9) read as follows:

> … a common measure, the notion of any correction beyond what we have instruments and art to make, is a mere fiction of the mind, and useless as well as incomprehensible. But tho' this standard be only imaginary, the fiction however is very natural […] This appears very conspicuously with regard to time; where tho' 'tis evident we have no exact method of determining the proportions of parts, not even so exact as in extension, yet the various corrections of our measures, and their different degrees of exactness, have given us an obscure and implicit notion of a perfect and entire equality.

Hume notes that our intuitive notion of accuracy is based on an imagined universal standard, "a common measure." It is however a fiction. Although from the folk (Newtonian) perspective clock adjustment is about imitating the one true flow of time, all we have for our temporal measures is the technology available to us. This will never yield "a perfect and entire equality."

About one hundred and fifty years later, Mach (1893/2013, 224) argued in his *Science of Mechanics* that Newton's absolute time is a trifling metaphysical concept that "can be measured by comparison with no motion." The uniform ticking of a clock means that it ticks at the same pace in relation to another clock. Whether the sequence of ticks in itself is uniform makes as much sense as asking whether a motion in itself is uniform. Time relates to change. Under relationism, change is a comparative phenomenon. In assessing the standards of time-measurement, one periodic process is compared to another one, not to putative time itself:

> When the physicist wishes to determine a period of time, he applies, as his standards of measurement, identical processes or processes assumed to be identical, such as vibrations of a pendulum, the rotations of earth, etc. (Mach 1886/1914, 349).

Although the relationist arguments of Hume and Mach already challenge the notion of absolute standard of temporal measurement,[5] Carnap went on to detail metrical conventionalism about clocks in his *Philosophical Foundations of Physics*. This work was published in 1966; it was however based on previous courses, in particular one taught at UCLA in 1958. In the book, Carnap starts by noting how clocks are simply instruments for creating periodic processes. Then he distinguishes between weak and strong definitions of periodicity. In the weak sense, something is periodic if it occurs repeatedly. A swinging pendulum, heartbeat, and Mr. Smith leaving his house are all periodic in the weak sense, as we are dealing with processes that keep happening again and again. Provided that one phase of the process continues to repeat, we have a minimally periodic process.

The stronger definition of periodicity, in addition of satisfying the condition of repetition, takes it also to be "true that the intervals between successive occurrences of a certain phase are equal" (Carnap 1966, 80). Mr. Smith does not exit his house at completely equal intervals of time. He might leave off at different times on different days, and some days he might stay home. Properly constructed clocks are periodic in a stronger sense, that is, they instantiate equality between successive periods. Although Carnap initially approves such a distinction – he also allows there to be "an enormous difference between the two types of periodicity" (Carnap 1966, 80) – he declines that there is a correct basis for measuring time.

In the view of Carnap, there is an inherent circularity involved in defining periodicity in the strongest possible way. What we need first is the method for assessing equal temporal intervals. This method is derived from conventionally chosen rules. Consequently, equal intervals of time are equal because we have already established the equality of adjacent periods by means of our rules. The only way out of the vicious circle is

[5] As well as absolute time in general. It is well-known that Hume and Mach paved the way for special relativity (Norton 2004).

> by dispensing entirely with the requirement of periodicity in the strong sense. We are forced to abandon it, because we do not yet have a basis for recognizing it. We are in the position of a naïve physicist approaching the problem of measuring time without even the advantage of prescientific notions of equal time intervals. Having no basis whatever for time measurement, he is seeking an observable periodic process in nature that will furnish such a basis. Since he has no way of measuring time intervals, he has no way of discovering whether a particular process is or is not periodic in the strong sense (Carnap 1966, 80–1).

Carnap introduces what he calls a three-fold schema as a method for measuring time. The schema includes rules for units, additivity, and equality. This can be clarified with the following sketch:

[INSERT FIGURE HERE]

Figure 1. Carnap's (1966, 81) figure for temporal measurement schema.

Units are the durations of any of the periods in the schema, A–B, B–C, C–D, or D–E. If the lengths of the durations are so that A–B is 1, A–C is 2, A–D is 3 and A–E is 4, we are ready to count the number of times that the unit period occurs. The rule of equality tells us that two intervals of time are equal in the case they contain the same number of periods. Hence in Carnap's schema, the temporal distance from A to C equals the temporal distance from C to E as both A–C and C–E contain two periods.

Carnap adds that periodic processes are commensurate. Say there is a shorter pendulum that produces a period *P*, and a longer one that produces a period *P*'. Although individual oscillations of the two pendulums are not equal, in large enough numbers one can establish equality among the periods. Ten swings of the shorter pendulum coincide with roughly six swings of the longer one. If it is the case that after ten oscillations of the shorter pendulum have occurred, the seventh oscillation of the longer one has already begun, we have not yet established equality. This is simply because fractions won't produce that equality. In case hundred successive occurrences of *P* matches with sixty-two successive occurrences of *P*', we have an equality. "It is a fact of nature," Carnap (1966, 83) has it,

> that there is a very large class of periodic processes that are equivalent to each other in this sense. This is not something we could know a priori. We discover it by observing the world. We cannot say that these equivalent processes are strongly periodic, but we can compare any two of them and find that they are equivalent.

This is all there is to the strong notion of periodicity. In such strong cases, which are the basis of our time-measurement practices and bureaucracy, clock smiths and engineers can compare different periodic processes across various phenomena. In weaker cases, like that of Mr. Smith's pulse, we could compare his heartbeat to a very limited class of periods, like some other rhythms within his body. The pulse of his wrist equals the one on his neck, but not pendulum swings, earth's rotation around its axis, or regular atomic vibrations. Clocks based on periodic quantum phenomena provide a coherent temporal measure of seconds, hours, days, and years. This could not be achieved with some individual's heartbeat.

The main difference between stronger and weaker periodic processes is not that the former provides a true description of time and the latter a false one. Instead, the former yields a simpler and less complex description of nature. We could use Mr. Smith's pulse, construct a temporal unit system based on it and assess the frequency of consecutive noons. We would judge days to be infrequent. There would be nothing wrong with this judgment, as the basis of synchronization is up to us to choose. Choosing strong periodic process as the standard of clocks is beneficial in rendering our description of the laws of physics as simple as possible.

Carnap's conventionalism finds support from Grünbaum (1962), among others, who thought that time is metrically amorphous. Nowadays, Rovelli expresses sympathies toward conventionalist views. In accounting for Galileo's innovation of comparing pendulums and pulses, he notes that virtually all clocks are oscillation-based and today it is customary to check our pulses by comparing them to ticking of clocks. Then Rovelli (2004, 20) asks: "What is the actual meaning of the pendulum periods taking "equal time"? An equal amount of *t* lapses in any oscillation: how do we know this, if we can access *t* only via another pendulum?" External variables are measured against clocks; they do not imitate a putative absolute time.

In the absence of Newtonian time, all we have are relative measures of time. We have chosen the basic standard so to produce a simple and coherent system of measuring time. Clocks produce equal periods in the sense their periods are commensurate. Strong conventionalism denies that one clock could be more accurate than another, because the standard by which we assess the equality of subsequent periods is founded on our rules that originally define equality.

## 4. The shortcomings of strong conventionalism

Many arguments along conventionalist lines are agreeable. Congruence is to some degree dependent on our preferences because the notion of stability is idealized. The standard second, 9,192,631,770 periods, is both a definition of the duration of the second and a stipulation that a particular atomic frequency is uniform. There is extensive human idealization involved in judging that the periods of caesium's vibration are congruent. The standard clock should be at rest at zero degrees Kelvin, without any background fields influencing the transition (Tal 2016, 301). Moreover, there are roughly 200 standard clocks distributed around the globe, so we are dealing with some average measure of these clocks. Time dilation due to relative motion of a walking speed is already significant, not to mention elevation differences, like different floors of a building as well as local geology, tides, and magna shifts (Gibbs 2002, 93). In more mundane circumstances, there are many factors to be considered that make clocks drift: heat, dust, the effect of dropping the clock, immersion in liquids etc.. In terms of accuracy, there is also a theoretical limit. Clocks cannot measure events that last less than how long it takes for electromagnetic signals to traverse between the component parts of the clock that generate the sequence of ticks.

Based on these observations, it should be admitted that conventionalism sheds light on numerous important issues about temporal measurement. In this sense, it is an astute theory of clocks. The strong version of conventionalism, however, does not present a cogent perspective on clock measurement. Below three reasons to think so are provided.

## 4.1 Periodicity differs in terms of regularity

Despite there are multiple conventional elements about clock synchronization, strong conventionalism implies eventually that there is no difference between different periodicities in terms of their regularity. It might be right that there is no natural, objective difference between periodicity in the weak and the strong sense. Does this mean there is no difference between complete randomness and highly regular processes what-so-ever?

Carnap himself hints that there are more and less accurate clocks. In explaining how the periodic processes of clocks can be created in various ways, he notes that the previous basis, the periodic rotation of the earth, was discarded because "the rate of the earth's spin is changing slightly", so "in 1956 an international agreement was reached to base units of time on the periodic movement of the earth around the sun in one particular year." This was subsequently "abandoned in 1964 so that still greater precision could be obtained by basing the second on the periodic vibration rate of the cesium atom" (Carnap 1966, 79). How is this not describing a method of acquiring a better clock? He does implicitly allow that some periods are changing. What does it mean to say that a period changes? With respect to what this change occurs? If a planet's spin changes, however slightly, that means there is some real or ideal standard to which the period of the planet is compared to. Carnap himself suggests that planetary periods change more and quantum periodicities less. It is in virtue of decreasing change in the periodicity, in his own words, "that still greater precision could be obtained". That is the way to improve the accuracy of time measurement.

Strong conventionalism should deny the significance of lack of perturbation to clock accuracy. "What is the warrant," Grünbaum (1962, 408) asks, first in reference to spatial dimension, "that a solid rod remains *rigid* under transport in a spatial region *free* from inhomogeneous thermal, elastic, electromagnetic and other "deforming" or "perturbational" influences?" In relation to temporal dimension: "What are the grounds for asserting that a clock which is not *perturbed* in the sense just specified is *isochronous*?" (Grünbaum 1962, 408). If some clocks are better than others in terms of accuracy, then the answer to Grünbaum's challenge is the higher degree of regularity. Dowden (2025) suggests that a better standard clock "will tick *regularly* in the sense that all periods between adjacent ticks are sufficiently congruent, that is, the same duration." Clocks based on the beats of someone's heart are not based on as regular processes as atomic clocks because they get out of synchrony much more easily. Although there are idealizing conditions even in the case of more refined atomic standards, these standards themselves are much more regular in their operation than most of the other alternatives that are available for clockmakers. The more regular periods are less disturbed by external factors.

It should be admitted that the previous reasoning hinges on an intuition about the difference between regularity and randomness. Although Carnap (1966, 80) shares this intuition (by his own words, he admits that there is “an enormous difference between the two types of periodicity,” the strong and the weak, thus contradicting his ultimate position of strong conventionalism), this in itself does not yet establish a non-conventional notion regularity. How could we ever know how some natural processes are regular as opposed to random? How to even know whether one process is more regular than another one, that is, how to even establish a difference of degree? The answer is that highly regular processes do not follow the rules set up by humans. The regularity of atomic vibrations is not solely based on human conventions and definitions. We humans cannot adjust the rate of periodicity of quantum hyperfine transitions, whereas we can easily change our pulse by varying ways. It is also notable that we cannot adjust many other natural periodic processes, like half-life of elementary particles. Muons hit the ground of the earth in the numbers as they do because they are subject to time dilation. This is an ontological point: muons exist at the ground the way they do because of time dilation. The existence of muons at the earth does not follow from human conventions. Half-life decays are not exactly clocks because they do not count ticks or display time. They are nevertheless periodic in their operations, just like the standard atomic clocks are. Even though periodicity relates in some degree to our conventions, just like other foundational concepts in time measurement like equality, stability, and uniformity, periodicity is not merely conventional. As periodicity is the basis of our clocks, the basis of our clocks is not completely conventional, either.

One might still remain unsure about how convincing the counter-arguments in this subsection are. Is it not still a stipulation to say that decay processes happen with a certain regularity? Additionally, what is the difference between actively constructing or adjusting a process and then stipulating it as a standard of periodicity, versus merely observing a process and stipulating it as a standard of periodicity? It must be admitted that there is no non-circular way to indicate that one process is more regular and temporally more equal than another. There is *some* convention involved. Yet periodicities that are the basis of making rules about equality and regularity do not hinge *merely* on convention. Even though the measurement processes involved in experimenting with muons hitting the ground are not convention-free, the arrival of muons at the detectors on Earth's surface do not follow from human actions. They are due to cosmic ray and upper atmosphere particle interactions and the specific properties of the elementary particles. We humans certainly did not tweak the basic properties of any lepton. Those properties determine how long the particles in question last. Muon's mean-life is not principally different from, say, electron's mass: neither follow from human conventions. In muons' case, there has to be some non-conventional temporal element related to it. Such determinate elementary particle properties differentiate natural regularities observed and measured in nature from artificially created and constructed processes.

Strong conventionalism posits that regularities arise from our conventions. Although conventionalism does not alter the ontological status of periodicities, it still challenges the intuition that atomic periodicities are intrinsically more regular than a person's heartbeat. Carnap urges us to dispense "entirely with the requirement of periodicity in the strong sense" because it impossible to "know that a process is periodic in the strong sense unless we already have a method for determining equal intervals of time!" (Carnap 1966, 80–1). This can be seen as a questionable account of the standards of synchrony, as it makes the standard entirely depend on the rules we have set up.

### 4.2 Accurate clocks measure the parts of their timelike worldlines

To quote from Carnap (1966, 79): "Every clock is simply an instrument for creating a periodic process." This account does not explain what clocks actually measure. A clock is not only a device that imitates periodicities. Properly constructed clocks track a real physical quantity. Ideally clocks measure parts of their own trajectories in spacetime. Different paths in spacetime yield different clock readings. According to Maudlin's (2012, 76) analogy, clocks are like odometer readings. If two cars have initially the same mileage, and then they go on and drive different spatial paths, their readings will not match at the rendezvous. Odometer measures are one-dimensional scalar quantities, whereas relativistic observers traverse in four-dimensional spacetime. The proper time, $\tau$, is the integral along the observer's worldline $\tau = \int d\tau$, where $d\tau = \sqrt{c^2dt^2 - dx^2 - dy^2 - dz^2}/c$ in which $t$ is the coordinate time, $x$, $y$, and $z$ the three spatial coordinates, and $c$ the speed of electromagnetic signals. Proper time is an invariant quantity. Unlike coordinate time, its value is independent of the frame of reference used to calculate it (Arthur 2006, 141). The coordinate time cannot be measured; a clock measures approximately its own proper time.

One definition of clock accuracy is that an accurate clock measures the path it travels along its trajectory in spacetime. The measure of the path is the length of the worldline between an initial event, $E_i$, and a final event, $E_f$, so the proper time is $\int_{E_i}^{E_f} d\tau$. In case the clock tracks the quantity of proper time correctly, it has measured the timelike path in the interval $E_f - E_i$. We may idealize and assume that ticks of a clock are instantaneous. The amount of time elapsed between adjacent ticks should be the same amount of time. The ratios of differences in the clock's way of assigning numbers to events should be proportional to the ratios of the interval lengths of the parts of the worldline. Provided an ideal clock assigns number 1 to an event $A$, number 3 to an event $B$ and number 7 to an event $C$ along its worldline, then the ratio of the length of segment $\overline{AB}$ to the length of segment $\overline{BC}$ should be 1/2 (Maudlin 2012, 108).[6]

If we adopt a scientifically realist position on the orthodox interpretation of special relativity,[7] time dilation is a real phenomenon. Different observers truly age at different rates in comparison to each other. Accurate clocks ideally measure the length of the interval the observer travels in spacetime, so different trajectories yield different proper times which are reflected on the different readings of adequately working clocks. Same must hold for ideally accurate odometers as they gauge different mileages for varying spatial distances. In both cases, the elapsed temporal and spatial lengths are calculable and practically testable with flying out atomic clocks and driving cars.

There is nevertheless a difference between clock and odometer measurements. The former are cyclical in a way that the latter are not. In case a clock recognizes no difference between a.m. and p.m., it restarts every 12 hours. In case a clock recognizes the difference between a.m. and p.m., it restarts every 24 hours. Your wristwatch does not tell you how long it has been traversing in spacetime. In contrast, the odometer tells you how long you have driven your car. Whereas an odometer measures the length of its path, the interval in question, the clock does not have resources to register its entire path due to its repetitive structure. There might be other kinds of time-measuring devices that measure the complete interval, devices which tick from its creation to its destruction, counting all the ticks on its path. They would not be clocks, if clocks are strictly defined by means of periodicities.[8]

Accuracy is not merely a matter of keeping clocks in synchrony with the conventionally chosen periodic standard, as strong conventionalism indicates. Ideally accuracy is about the measure of proper time, the part of the interval of the clock's own trajectory in spacetime. Strong conventionalism does not explicate this, because it does not consider the quantity that clocks aim to measure.

---

[6] The clock hypothesis is further developed by Fletcher (2013).

[7] This of course excludes obsolete Lorentzian ether-based accounts.

[8] Time-measurement devices drift, and so they need to be synchronized with the aid of a standard clock that is based on a conventionally chosen standard. That is why non-periodic time-measuring devices are not clocks.

There is, however, a plausible avenue for the conventionalist to counter the criticism presented in this subsection. The supporter of the conventionalist stance might argue that the theory of relativity incorporates a predetermined convention for defining proper time, even if it is not explicitly acknowledged. Proper time within the framework of relativity is typically defined through the periodic oscillations of an elementary light clock. This essentially ties the definition of proper time to a particular physical process deemed standard within that convention. An entirely different periodic physical process could have been chosen, resulting in an alternate convention and a corresponding redefinition of proper time. In essence, the choice of what constitutes the periodic standard reflects a conventional decision, not an unavoidable truth of the natural world.

As with other parts of this article, the aim is not to argue that strong conventionalism is entirely incorrect. Instead, the goal is to demonstrate that this represents yet another area where the strong version of conventionalism encounters a significant challenge. Since clocks are instruments that ideally measure something external to themselves, there must exist a specific quantity that they track. This targeted quantity, as will be discussed in the subsection below through a comparison with measurement units, is not a matter of convention.

An important question arises in this context. While neither the clock itself nor any subject actively generates the path the clock follows through spacetime, or the proper time along that path, is the way we segment that path into repeated, equal intervals truly reflective of an intrinsic aspect of the trajectory? Or is this method of division merely a matter of human convention? For instance, do we designate a cycle of 24 hours because it corresponds to some naturally occurring segmentation of the clock's proper time, or simply because it is convenient and aligns with the Earth's rotation? Although a committed conventionalist would not assert that a clock determines its own spacetime trajectory, they might reasonably maintain that the portions of that trajectory we track, such as second after second, are themselves products of human-defined conventions. A clock tracks a conventionally determined part of its trajectory.

It is true that, in the aforementioned sense, the partitioning is up to us. A 24-hour day is convenient for us earth inhabitants, and that is what conventionalism reasonably suggests. However, the ratios and magnitudes themselves are not subject to our discretion. For instance, the ratio of $3 - 1$ to $7 - 3$ is $1/2$, as explained in Maudlin's earlier example. One might, of course, adopt a conventionalist stance regarding such mathematical truths. However, I will not delve into a discussion on the philosophy of mathematics, as it falls outside the scope of this paper. Instead, I wish to focus on magnitudes and the invariant nature of proper time.

When comparing two trajectories that start at the same point and come back together, their proper times won't match in case their paths are different. The maximum amount of proper time is elapsed along the straight interval, $\Delta\tau_{STRAIGHT} = \Delta t$. A non-straight interval has lesser proper time elapsed, $\Delta\tau_{NON-STRAIGHT} = \sqrt{1 - v^2}\Delta t < \Delta t$. We may choose the cycle of repetition as we like, but we cannot choose $\Delta\tau_{STRAIGHT} > \Delta\tau_{NON-STRAIGHT}$. The non-equality of these two paths is not conventional, although the specific cyclical processes we use to track those magnitudes are. There is a real quantity behind the choice of length of the repetitive process, the length of the part of the worldline.

The quantitative values associated with repeating motions are not conventional. We can precisely count how many times the second hand completes a full rotation past the top of the clock, just as we can tally the exact number of wheel rotations recorded by an odometer. These values are relative to specific paths, yet there is only one correct answer regarding how many times the second hand or wheel rotates around its axis. These numbers are not matters of convention.

The issue for strong conventionalism lies in its failure to provide an explanation for how the accuracy of clock measurement is rooted in the reference of clocks to something external. Proper time is an invariant and objective characteristic of four-dimensional spacetime, something existing independently of our social conventions. Trajectories in spacetime traversed by clocks are not conventional the same as trajectories in one spatial dimension traversed by cars are not conventional. In both cases we are dealing with a definite quantity.

### 4.3 Temporal units ≠ temporal quantities

We may discern at least one temporal measure which is plainly objective. A shorter duration within a longer one is shorter than the longer. If writing this paper takes in total three months, and writing one of its sections within that period takes three weeks, that shorter period is contained in the longer one and shorter than the longer one. We cannot make any kind of convention about the interconnections of such lengths. Units of time, for their part, are doubtlessly conventional. We can use jiffys, seconds or tithis and choose the base unit as we please. Units are neither logical necessities nor discovered in nature. They are rather up to the specific culture that has adopted a contingent way of assigning units to their temporal measures.

Although it is up to our whim that we have decided to use seconds as the basis of the unit of time, clocks measure a physical quantity, much like scales are calibrated to reveal the total quantity of matter of an object and thermometer readings correspond to the average kinetic energy of a set of molecules. Consider comparing two masses and two temperatures. A big adult stands on one scale, and a small child on a second scale. The former scale gives a reading of 10kg and the latter a reading of 100kg. Although all measurements involve error estimates, we can safely judge that these scales report blatantly wrong numbers. Say then there are two thermometers. The first is stuck in a bucket of ice, the latter in boiling water. The former shows 100C° and the latter 0C°. Again, we may conclude that there is something wrong with the mechanism or calibration of these measuring instruments. If an atomic clock transported around the world in a high-speed jet does not suffer time dilation in comparison to a previously synchronized stationary clock, we should declare that the clocks in these cases are in error. Either one or both of the clocks would not have tracked the quantity they are supposed to track, namely proper time, the measure of the segment of their worldlines.

One might argue that this point – while the selection of temporal units is a matter of convention, this does not imply that the nature of temporal quantities themselves is conventional – holds little relevance to strong conventionalism. It does not base its position on any specific claim about the relation between temporal units and temporal quantities. Instead, strong conventionalism focuses primarily on social agreements, linguistic conventions, and creation of rules.

A relevant critique of strong conventionalism is that it lacks the resources to acknowledge that a successful measurement reflects an objective feature of reality. It tacitly implies that measuring devices create what they measure. Carnap does not explicitly make such a claim, but he does implicitly espouse a position like this, as Carnap (1966, 79) claims that each "clock is simply an instrument for creating a periodic process." This calls into question the credibility of strong conventionalism as a framework for understanding temporal measurement; strong conventionalism does not explain that measurement connects with something that exists independently of the measuring device itself.[9]

[9] The issue of creating a quantity is problematic for strong constructivism about clocks, something Latour (1987, Section "Tied in by a few metrological chains") and Schaffer (1992) arguably have in mind. In the interpretation of Tal (2016, 316), social constructivists are committed to the following: "quantitative scientific claims attain universal validity not by virtue of any pre-existing state of the world, but by virtue of the continued efforts of metrologists who transform parts of the world until they reproduce desired quantitative relations". If this interpretation is correct, clocks would not track any pre-existing quantity. This would be a very radical view of measurement. What would a scale measure under such strong constructivism? The scale would have to construct the mass it purports to measure, essentially creating atoms out of nothing!

If there is a quantity, then that quantity can be measured, and there is more or less of that quantity. That makes a comparison possible. Quantities can be thought of as primary in the sense that one does not even need units for there to be quantities. Considering clock time as the path in spacetime, different observers yield different values in comparison to each other. If the first observer is stationary, $v_1 = 0$, and the second travels with half the speed of light, $v_2 = 1/2c$, then time passes more quickly for the first observer in comparison to the second and time passes more slowly for the second observer in comparison to the first.[10] Let's formulate the special relativistic metric equation, $c^2 d\tau^2 = c^2 dt^2 - dr^2$, in which $\tau$ is the proper time, $t$ the coordinate time and $r$ the spatial dimension, into finite quantities, $c^2 \Delta\tau^2 = c^2 \Delta t^2 - \Delta r^2$. We can easily calculate how it yields $\Delta\tau_1 = \Delta t_1$ for the first observer and $\Delta\tau_2 = \sqrt{3/4}\Delta t_2 \approx 0.87\Delta t_2$ for the second. This means about 115 precent of time passes for the first observer in relation to the second observer, and about 87 percent of time passes for the second observer in relation to the first.[11] Likewise, we could say that person number one, who is 1.9m tall, is about 115 percent of the height of person number two, who is 1.65m tall, and person number two is about 87 percent of the height of the first person. In both cases, we are dealing with quantities that are measurable, and there are more or less of them.

Quantities are clearly different from bare units. The remarkable difference between quantities and units is that the former are not conventional whereas the latter are. Proper time is a real quantity – its magnitude is not subject to our discretion, unlike the unit system we use. A robust theory of clock measurement should be capable of explaining the distinction between temporal units and quantities, particularly in terms of conventionality versus non-conventionality. Strong conventionalism about clocks lacks this robustness.

[10] One might object that this comparison gets us to the so-called twins paradox. Time dilation is a symmetric effect: each observer finds the clocks of the other lagging. Given the different paths in spacetime explanation, the first and the second observers' cases are not symmetrical. Considering proper time, $-dx^2 - dy^2 - dz^2$ yields 0 for the first observer but $\neq 0$ for the second one. From this perspective, there is nothing paradoxical about different clock readings.
[11] This evidently relates to the challenge of whether there is passage of time if there is no real dimension of passage. Price (1996, 13) notes $1s/s = 1$ yields a mere dimensionless 1 and no unit. I shall not enlarge on this point – whether there is genuine passage of time in relativistic spacetime – further as it would force me to deviate too much from the topic of this article. The important point here is that even if proper time does not have an adequate unit, there is evidence for its quantitative existence.

## 5. Conclusion

This article explored the issue of clock measurement by centering around strong conventionalism. As the standard of synchrony is a conventional matter, conventionalism implies that no standard is true or false, but some standards provide more simple and coherent measurement systems. For its part, a realistically inclined position indicates that the ticks and recurring periods are ideally isochronous in the sense that they track a physical quantity, namely the parts of the clock's timelike worldline. Although conventionalism is in many ways a valid way of understanding temporal measurement, it was argued that in its most extreme formulation it cannot properly account for three interrelated aspects: 1) There are differences in the regularities of periodic processes that are salient for improving temporal measurement; 2) Clocks ideally measure a physical quantity, its timelike curve in spacetime; 3) That quantity measured is not conventional, unlike the temporal units we happen to use. These critical arguments do not show that conventionalism about clocks is erroneous, yet alone that conventionalism would be a completely wrong theory about measurement in general. Based on these critical arguments, we may nevertheless conclude that strong conventionalism about clocks is not a convincing position about temporal measurement.

## References

Amiriara, Hassan (2021) “On the possibility of non-eternalism without absolute simultaneity.” *Synthese* 199: 5885–98.

Arthur, Richard (2006) “Minkowski Spacetime and the Dimensions of the Present.” In Dieks, D. (ed.) *The Ontology of Spacetime*, 129–56. Amsterdam: Elsevier.

Balashov, Yuri and Michel Janssen (2003) “Presentism and Relativity.” *British Journal for the Philosophy of Science* 54: 327–46.

Blackburn, Simon (1996) *The Oxford Dictionary of Philosophy*. Oxford: Oxford University Press.

Carnap, Rudolf (1966) *Philosophical Foundations of Physics*. Martin Gardner (ed.). New York: Basic Books.

Deng, Natalja (2019) “One Thing After Another: Why the Passage of Time Is Not an Illusion.” In Arstila et al. (eds.), *The Illusions of Time*, 3–16. Cham: Palgrave-MacMillan.

Dowden, Bradley (2025) “Time.” *The Internet Encyclopedia of Philosophy*, ISSN 2161-0002, https://iep.utm.edu/time/.

Einstein, Albert (1905) “On the electrodynamics of moving bodies.” In *The collected papers of Albert Einstien Volume 2: The Swiss years: Writings, 1900–1909* (English translation supplement), 140–71. Translated by Anna Beck.

Farr, Matt (2022) “Conventionalism about time direction.” *Synthese* 200: 23, 1–21.

Fazekas, Katherine (2016) “Special Relativity, Multiple B-series, and the Passage of Time.” *American Philosophical Quarterly* 53 (3): 215–29.

Fletcher, Samuel C. (2013) “Light Clocks and the Clock Hypothesis.” *Foundations of Physics* 43: 1369–83.

Gibbs, Wayt (2002) “Ultimate Clocks.” *Scientific American*, September 2002, 86–93.

Craig, W. L. (2001) *Time and the metaphysics of relativity*. Dordrecht: Kluwer.

Grünbaum, Adolph (1962) “Geometry, Chronometry, and Empiricism.” In *Scientific explanation, space and time*, 405–526. Minnesota studies in the philosophy of science, Volume 3.

Gödel, Kurt (1949) “A remark about the relationship between relativity theory and idealistic philosophy.” In P.A. Schlipp (ed.), *Albert Einstein: Philosopher-scientist*, 555–62. La Salle, IL: Open Court.

Hume, David (1739–40/2007) *A Treatise of Human Nature*. Edited by David Fate Norton and Mary Jane Norton. Oxford: Clarendon Press.

Latour, Bruno (1987) *Science in Action*. Cambridge, MA: Harvard University Press.

Leininger, Lisa (2021) "Temporal B-Coming : Passage without Presentness." *Australasian Journal of Philosophy* 99 (1): 130–47.

Janis, Allen (2018) "Conventionality of Simultaneity." *The Stanford Encyclopedia of Philosophy*. Edited by Edward N. Zalta. https://plato.stanford.edu/entries/spacetime-convensimul/.

Johnston, Hamish (2021) "Three top atomic clocks are compared with record accuracy." Metrology research update, *Physics World*, March 26, 2021, https://physicsworld.com/a/three-top-atomic-clocks-are-compared-with-record-accuracy/.

Mach, Ernst (1893/2013) *The Science of Mechanics. A Critical and Historical Exposition of its Principles*. Translated by Thomas J. McCormack. New York: Cambridge University Press.

Mach, Ernst (1886/1914) *The Analysis of Sensations*. Translated by C. M. Williams and Sydney Waterlow. Chicago and London: Open Court.

Maudlin, Tim (2012) *Philosophy of Physics*: Space and Time. Princeton: Princeton University Press.

Norton, John D. (2004) "How Hume and Mach helped Einstein find special relativity." In Friedman, Michael, Mary Domski and Michael Dickson (eds.), *Discourse on a New Method: Reinvigorating the Marriage of History and Philosophy of Science*, 359–86. Open Court.

Peterson, Daniel and Michael David Silberstein (2010) “Relativity of Simultaneity and Eternalism: In Defense of the Block Universe.” In V. Petkov (ed.), *Space, Time, and Spacetime: Physical and Philosophical Implications of Minkowski’s Unification of Space and Time*, 209–37. Heidelberg: Springer.

Rescorla, Michael (2019) “Convention.” *The Stanford Encyclopedia of Philosophy*. Edited by Edward N. Zalta. https://plato.stanford.edu/entries/convention/.

Price, Huw (1996) *Time’s Arrow and Archimedes Point*. New York: Oxford University Press.

Rovelli, Carlo (2004) *Quantum Gravity*. Cambridge University Press.

Rovelli, Carlo (2023) “How Oriented Causation is Rooted into Thermodynamics.” *Philosophy of Physics* 1 (1): 11, 1–14. https://doi. org/10.31389/pop.46

Saunders, Simon (2002) “How Relativity Contradicts Presentism.” *Royal Institute of Philosophy Supplement* 50: 277–92.

Schaffer, S (1992) “Late Victorian Metrology and Its Instrumentation: A Manufactory of Ohms”, in R. Bud and S. E. Cozzens (eds), *Invisible Connections: Instruments, Institutions, and Science*. Washington, DC: SPIE Optical Engineering Press, pp. 23–56.

Slavov, Matias (2021) “Hume’s Thoroughly Relationist Ontology of Time.” *Metaphysica* 22 (2): 1–16.

Slavov, Matias (2022) *Relational Passage of Time*. New York: Routledge.

Swinburne, R. (2008) “Cosmic simultaneity.” In W. L. Craig and Q. Smith (eds.), *Einstein, relativity and absolute simultaneity*, 224–61. London: Routledge.

Tal, Eran (2016) “Making Time: A Study in the Epistemology of Measurement.” *The British Journal for the Philosophy of Science* 67: 297–335.